# Understanding Modal Interactions in Non-classically Damped Linear Oscillators with Closely Spaced Modes

Luis M. Baldelomar Pinto[1], Alireza Mojahed[2], Sobhan Mohammadi[3], Keegan J. Moore[3], Lawrence A. Bergman[1], Alexander F. Vakakis[1]

[1] University of Illinois Urbana-Champaign, Urbana, IL 61801
[2] 150 Glover Avenue, Norwalk, CT 06850
[3] Georgia Institute of Technology, Atlanta, GA 30332

**Abstract**

Linear discrete coupled oscillators with closely spaced modes have been a topic of intense study since they appear often in engineering practice, e.g., in turbomachinery applications, and jointed structures with symmetries. When these systems are classically viscously damped, their vibration modes cannot interact (i.e. are not coupled and, hence, cannot exchange energy), and each vibration mode represents an independent two-dimensional (2D) invariant manifold in phase space. However, this does not hold in non-classically (non-proportionally) damped systems, for which, it is well known to possess complex vibration modes; for such systems closely spaced vibration modes do interact, and represent higher-dimensional (that is, higher than 2D) invariant manifolds in phase space. This work addresses this classical vibration problem from a new perspective, focusing on the physics (dynamics) of modal interactions in non-classically damped, linear systems with closely spaced modes. Considering the simplest representative example in the form of an impulsively excited two-degree-of-freedom (two-DOF) system, we show that there is a single parameter defined as a coupling versus damping non-proportionality ratio, that separates two different dynamical regimes. Based on complexification-averaging analysis, we show that, below the critical value of this parameter, the system response possesses two distinct dissipation rates but only one frequency of oscillation; as a result, energy is slowly exchanged between modes in a single beat phenomenon. However, above the critical parameter value, the response has a single dissipation rate but two distinct oscillation frequencies; this yields an infinity of beat phenomena as energy is interchanged at a faster rate between modes. Our analytical predictions are fully validated by experimental measurements. Our findings highlight the physics of modal interactions in coupled oscillators and provide a framework for system identification and reduced-order modeling of systems with closely spaced modes.

# 1. Introduction

Linear coupled oscillators with closely spaced modes have been studied extensively since they are encountered often in engineering practice. As examples, we mention bladed disk assemblies in turbomachinery where there are numerous families of multiple closely spaced vibration modes, and multi-component structures possessing weak substructure coupling and/or spatial symmetries (Ewins, 2000; Garcia-Fernandez, 2022). Experimental system identification and reduced-order modeling of practical systems with closely spaced modes is typically performed in the frequency domain, and faces distinct challenges related to modal interactions, energy exchanges between modes and varying relative phase differences between modal responses. Hence, this is an area of active research which is still not fully explored, e.g., (Ewins, 2000; Tan et al., 2008; Kordkheili et al., 2018), especially when additional complications arise, e.g., due to nonlinear hysteresis effects induced by mechanical joints.

A main source of complication in systems with closely spaced modes has to do with their spatial damping distribution (typically viscous damping); indeed, in coupled oscillators with so-called *proportional damping distribution*, all modes are real (i.e., their mode shapes are real-valued) and uncoupled, so no energy exchanges can occur between them; this is the case of *classically damped coupled oscillators* (Caughey, 1960; Caughey and O'Kelly, 1965; Phani, 2003), where each mode is represented by an independent two-dimensional (2D) invariant manifold in phase space (these are the *uncoupled modal oscillators* of linear vibration theory). On the contrary, for non-proportional damping distribution, the vibration modes are coupled and complex (Ewins, 2000; Krack et al., 2016), so modal interactions occur during which modal energy exchanges are realized; in this case the system of coupled oscillators is said to be *non-classically damped*, and the vibration modes take place in high dimensional invariant manifolds (i.e., with dimensions >2) in phase space. Ma and co-workers (Ma et al., 2009; 2010) and (Salsa and Ma, 2021) developed a decoupling algorithm for computing independent (uncoupled) "damped modes" in a general class of non-classically damped oscillatory and non-oscillatory systems by means of time-varying coordinate transformations.

The aim of this study is to bring physical insight into the modal interactions that take place in non-classically damped coupled oscillators with closely spaced modes. Whereas solving the associated eigenvalue problems and formally computing the (complex) vibration modes or the "damped modes" of such systems is rather straightforward, a physical interpretation of the results in terms of modal interactions and modal energy exchanges, and their dependence on the system parameters is not well understood. Considering as a representative example an impulsively excited two-DOF symmetric system of weakly coupled and weakly damped oscillators, we employ the method of Complexification-Averaging to obtain leading order approximations for the modal responses. Perhaps surprisingly, we find that a single coupling to damping non-proportionality ratio governs two distinct dynamical regimes of the response. Below a critical value of this ratio, the dynamics is governed by two distinct dissipation rates but only one (fast) frequency of oscillation, and energy is slowly exchanged between the two modes of the system in a single beat phenomenon. However, above the critical value, the response has a single dissipation rate but two distinct closely spaced (fast) oscillation frequencies, which due to beating gives rise to a slow frequency, and an infinity of beat phenomena with energy being exchanged between modes at a fast rate. Our analytical predictions are fully validated by experimental measurements. Our findings elucidate the physics of modal interactions in coupled oscillators and provide a framework for system identification and reduced-order modeling of systems with closely spaced modes.

## 2. Complexification-averaging (CX-A) analysis of modal interactions

The two-DOF system under consideration is illustrated in Fig. 1. It consists of two coupled oscillators which have identical masses and grounding stiffnesses but differing grounding viscous dampers. Without loss of generality, we normalize the masses $m_i$ and grounding stiffnesses $\beta_i$, according to $\beta_1 = \beta_2 = m_1 = m_2 = 1$, and assume *weak coupling and weak damping*, $\beta, \lambda_1, \lambda_2 \ll 1$, with the three parameters being of the same small order; moreover, the damping in the coupling element is eliminated by setting $\lambda = 0$. Lastly, we note that the configuration symmetry of the system can be perturbed only when $\lambda_2 \neq \lambda_1$, which renders this system *non-classically damped*. Throughout this work we will make the assumption (without loss of generality) that $\lambda_2 > \lambda_1$, and denote the more lightly damped oscillator on the left as "oscillator 1", while the more heavily damped one on the right as "oscillator 2".

The governing equations of motion are given by,

$$\ddot{x}_1 + \lambda_1 \dot{x}_1 + x_1 + \beta(x_1 - x_2) = 0 \qquad (1a)$$
$$\ddot{x}_2 + \lambda_2 \dot{x}_2 + x_2 + \beta(x_2 - x_1) = 0 \qquad (1b)$$

where overdot indicates derivative with respect to time. With the system initially at rest we are interested in the response of the oscillators under impulse excitations applied to the two oscillators, which is equivalent to assuming the non-zero initial conditions $\dot{x}_1(0) = A_1$, $\dot{x}_2(0) = A_2$, $x_1(0) = x_2(0) = 0$, where $A_i, i = 1,2$ are the impulse intensities.

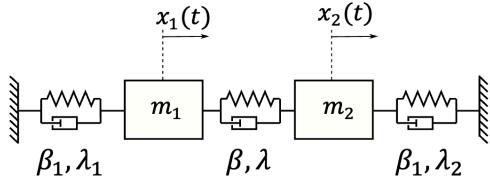

Figure 1. Non-classically damped system of coupled oscillators.

System (1) can also be expressed in terms of modal coordinates, $u_1 = (1/2)(x_1 + x_2)$ – the amplitude of the in-phase normal vibration mode, and $u_2 = (1/2)(x_1 - x_2)$ – the amplitude of the out-of-phase mode:

$$\ddot{u}_1 + (1/2)(\lambda_1 + \lambda_2)\dot{u}_1 + u_1 = (1/2)(\lambda_2 - \lambda_1)\dot{u}_2 \qquad (2a)$$
$$\ddot{u}_2 + (1/2)(\lambda_1 + \lambda_2)\dot{u}_2 + (1 + 2\beta)u_2 = (1/2)(\lambda_2 - \lambda_1)\dot{u}_1 \qquad (2b)$$

We note that for the case of *proportional damping*, $\lambda_2 = \lambda_1$, the in-phase and out-of-phase modes are uncoupled for any value of the coupling $\beta$, and no modal interactions can occur. For non-proportional damping, $\lambda_2 > \lambda_1$, however, the non-symmetric damping forces couple the two modes and modal interactions do occur. Hence, we focus exclusively on the latter case to study in detail the damping-induced modal interactions.

For weak coupling and damping, the oscillatory response of system (1) is expected to possess a single "fast" frequency equal to unity; this becomes apparent from the modal oscillators (2), since for $\beta, \lambda_1, \lambda_2 \ll 1$, the two natural frequencies are close to unity and the damping and coupling terms are small and of the same order. Hence, it is possible to analytically study the response of (1) through a slow/fast partition of the dynamics, in conjunction with the method of Complexification-Averaging (CX-A) first introduced by Manevitch (1999). To this end, we introduce the new complex variables $\psi_i = \dot{x}_i + jx_i$, $i = 1,2$, where $j = (-1)^{1/2}$, in terms of the responses and velocities of the two oscillators, and re-write system (1) in complex form,

$$\dot{\psi}_1 - j\psi_1 + \frac{\lambda_1}{2}(\psi_1 + \bar{\psi}_1) - \frac{j\beta}{2}(\psi_1 - \bar{\psi}_1 - \psi_2 + \bar{\psi}_2) = 0 \quad (2a)$$

$$\dot{\psi}_2 - j\psi_2 + \frac{\lambda_2}{2}(\psi_2 + \bar{\psi}_2) - \frac{j\beta}{2}(\psi_2 - \bar{\psi}_2 - \psi_1 + \bar{\psi}_1) = 0 \quad (2b)$$

where overbar denotes complex conjugate. Hence, the original second-order real equations of motion are expressed as two first-order complex equations. At this point we partition the complex responses in terms of "slow" complex envelopes $\phi_i$, and "fast" oscillations $e^{jt}$ at unit frequency, according to the *ansatz*, $\psi_i = \phi_i e^{jt}, i = 1,2$. It follows that the analytical results will be valid only in the range of parameters for which the above slow/fast partition is valid, i.e., as long as the responses are in the form of fast-varying oscillators with (nearly) unit frequency that are modulated by slow-varying envelopes.

Substituting the previous ansatz into the complex equations of motion (2) and averaging with respect to the fast frequency (this is equivalent to disregarding higher harmonic terms with fast frequencies higher than unity), we derive the following *complex slow flow dynamics*:

$$\dot{\phi}_1 + \frac{1}{2}(\lambda_1 - j\beta)\phi_1 + \frac{j\beta}{2}\phi_2 = 0 \quad (3a)$$

$$\dot{\phi}_2 + \frac{1}{2}(\lambda_2 - j\beta)\phi_2 + \frac{j\beta}{2}\phi_1 = 0 \quad (3b)$$

This set of equations governs the slow evolution of the envelopes of the oscillatory responses of the two oscillators, and, hence, the modal interactions that occur in the system with closely spaced modes, due to weak coupling and damping non-proportionality. For the previous impulse excitations, system (3) is solved subject to the initial conditions, $\phi_1(0) = A_1$, $\phi_2(0) = A_2$. This linear set of equations can be explicitly solved,

$$\phi_1(t) = e^{-\frac{\lambda_+ t}{2} + \frac{j\beta}{2} - \frac{j\lambda_-}{4}\sqrt{\gamma^2 - 4}} \left( \frac{\gamma A_2}{\sqrt{\gamma^2 - 4}} \left(1 - e^{j\frac{\lambda_-}{2}\sqrt{\gamma^2 - 4}}\right) \right. \\ \left. + \frac{jA_1}{\sqrt{\gamma^2 - 4}} \left(1 - e^{j\frac{\lambda_-}{2}\sqrt{\gamma^2 - 4}}\right) + \frac{A_1}{2}\left(1 + e^{j\frac{\lambda_-}{2}\sqrt{\gamma^2 - 4}}\right) \right) \quad (4a)$$

$$\phi_2(t) = e^{-\frac{\lambda_+ t}{2} + \frac{j\beta}{2} - \frac{j\lambda_-}{4}\sqrt{\gamma^2-4}} \left( \frac{\gamma A_1}{\sqrt{\gamma^2-4}} \left(1 - e^{j\frac{\lambda_-}{2}\sqrt{\gamma^2-4}}\right) \right.$$
$$\left. -\frac{jA_2}{\sqrt{\gamma^2-4}}\left(1 - e^{j\frac{\lambda_-}{2}\sqrt{\gamma^2-4}}\right) + \frac{A_2}{2}\left(1 + e^{j\frac{\lambda_-}{2}\sqrt{\gamma^2-4}}\right) \right) \quad (4b)$$

where three important parameters are defined as, $\gamma = \frac{4\beta}{\lambda_2-\lambda_1}, \lambda_+ = \frac{\lambda_2+\lambda_1}{2}, \lambda_- = \frac{\lambda_2-\lambda_1}{2}$. Specifically, $\gamma$ represents a coupling to damping non-proportionality ratio, $\lambda_+$ is the average of the two damping coefficients of the two oscillators, and $\lambda_- < \lambda_+$ characterizes the damping non-proportionality. Note that, even though we assumed that both the coupling and damping coefficients are small, the ratio $\gamma$ is an O(1) quantity. Moreover, from (4) one may deduce that the parameter $\gamma$ is the main parameter governing the modal interactions in the system of Fig. 1. Indeed, dynamics change qualitatively as $\gamma$ varies above or below the critical value $\gamma_{cr} = 2$: For $\gamma > \gamma_{cr}$, all exponentials containing the term $\sqrt{\gamma^2 - 4}$ are oscillatory, whereas, for $\gamma < \gamma_{cr}$ the same exponentials are exponentially decaying. Hence, *the critical value $\gamma_{cr} = 2$ defines two regimes of dynamics, namely, a regime for $\gamma < 2$ where exponential decay is predominant, and a regime for $\gamma > 2$ where the oscillatory behavior is predominant.*

These two distinct dynamical regimes can be explained more clearly when one employs the solution (4) to obtain analytical approximations for the physical responses of the two oscillators using the expressions, $x_i = (\phi_i e^{jt} - \overline{\phi}_i e^{-jt})/2j$, $i = 1,2$:

$$x_1(t) = A_1 \langle x_{11}(t) \rangle \sin(\omega_0 t) + A_2 \langle x_{12}(t) \rangle \cos(\omega_0 t) \quad (5a)$$
$$x_2(t) = A_1 \langle x_{12}(t) \rangle \cos(\omega_0 t) + A_2 \langle x_{22}(t) \rangle \sin(\omega_0 t) \quad (5b)$$

The slow envelopes $\langle x_{mn}(t) \rangle$ in (5) are given by,

$$\langle x_{11}(t) \rangle = \begin{cases} \frac{1}{2}(e^{-\sigma_1 t} + e^{-\sigma_2 t}) + \frac{1}{\sqrt{4-\gamma^2}}(e^{-\sigma_1 t} - e^{-\sigma_2 t}), & \gamma < 2 \\ e^{-\lambda_+ t/2}\left(\cos(\omega_d t) + \frac{2}{\sqrt{\gamma^2-4}}\sin(\omega_d t)\right), & \gamma > 2 \end{cases} \quad (6a)$$

$$\langle x_{12}(t) \rangle = \begin{cases} -\frac{1}{2}\frac{\gamma}{\sqrt{4-\gamma^2}}(e^{-\sigma_1 t} - e^{-\sigma_2 t}), & \gamma < 2 \\ -\frac{\gamma}{\sqrt{\gamma^2-4}}e^{-\lambda_+ t/2}\sin(\omega_d t), & \gamma > 2 \end{cases} \quad (6b)$$

$$\langle x_{22}(t) \rangle = \begin{cases} \frac{1}{2}(e^{-\sigma_1 t} + e^{-\sigma_2 t}) - \frac{1}{\sqrt{4-\gamma^2}}(e^{-\sigma_1 t} - e^{-\sigma_2 t}), & \gamma < 2 \\ e^{-\lambda_+ t/2}\left(\cos(\omega_d t) - \frac{2}{\sqrt{\gamma^2-4}}\sin(\omega_d t)\right), & \gamma > 2 \end{cases} \quad (6c)$$

and $\omega_0 = 1 + \frac{\beta}{2}, \sigma_{1,2} = \frac{\lambda_+}{2} \mp \frac{\lambda_-}{4}\sqrt{4-\gamma^2}, \omega_d = \frac{\lambda_-}{4}\sqrt{\gamma^2-4}$. The envelopes $\langle x_{mn}(t) \rangle$ analytically approximate the slow variation of the envelopes of the responses and represent the *slow dynamics*, whereas the harmonic functions with frequency $\omega_0$ represent the *fast dynamics*. In fact, one notes that for sufficiently weak coupling $\beta$ and damping coefficients $\lambda_1, \lambda_2$, the fast frequency is indeed close to unity, whereas the exponents $\sigma_1, \sigma_2$ and frequency $\omega_d$ characterizing the envelopes are small quantities. Hence, our previous *ansatz* regarding slow/fast partition of the dynamics holds, which underscores the validity of the analytical approximations to accurately model the impulsive responses of the non-classically damped system with closely spaced modes depicted in Fig. 1.

From the analytical results (5-6) the qualitative change in the dynamics resulting from a change in $\gamma$, becomes clear. *For $\gamma < 2$, the responses of the two oscillators possess two distinct dissipation rates, namely $\sigma_1$ and $\sigma_2$, but only one fast frequency $\omega_0 \approx 1$*; in this case the slow envelopes of the decaying oscillations are governed by pure exponentially decaying terms – see the upper expressions in (6a-c). However, *the dynamics change qualitatively for $\gamma > 2$ where the responses possess two distinct (but closely spaced) fast frequencies, namely $\omega_{d1} = \omega_0 - \omega_d$ and $\omega_{d2} = \omega_0 + \omega_d$, but only one dissipation rate $\lambda_+$*. It is interesting that the harmonic superposition of these two fast frequencies gives rise to a beat phenomenon, yielding a single fast frequency $\omega_0 \approx 1$ and a slow frequency $\omega_d$; hence, in this case the terms of the slowly decaying envelopes are exponentially decaying oscillations that generate recurrent energy exchanges between the oscillators in an infinite number of beat phenomena – see the expressions in (6a-c). We note that these findings validate the central assumption of the CX-A analysis, namely, that there is a single fast frequency close to unity present in the responses, which is satisfied at least up to moderate values of $\gamma$ (see discussion below).

In synopsis, we find that depending on the coupling to damping non-proportionality ratio, $\gamma$, there are two distinct regimes of oscillatory responses in the dynamics of the system of Fig. 1: For this ratio below the critical value, $\gamma < \gamma_{cr} = 2$ (i.e., when the coupling is relatively weak compared to the damping non-proportionality), the modal interaction is in the form of a single beat phenomenon, with energy being exchanged between oscillators at a relatively slow time scale; however, for $\gamma > \gamma_{cr} = 2$ (i.e., when the coupling is relatively strong compared to the damping non-proportionality), intense modal interactions occur in the form of an infinite number of beat phenomena with energy being continuously exchanged between modes at a relatively faster time scale.

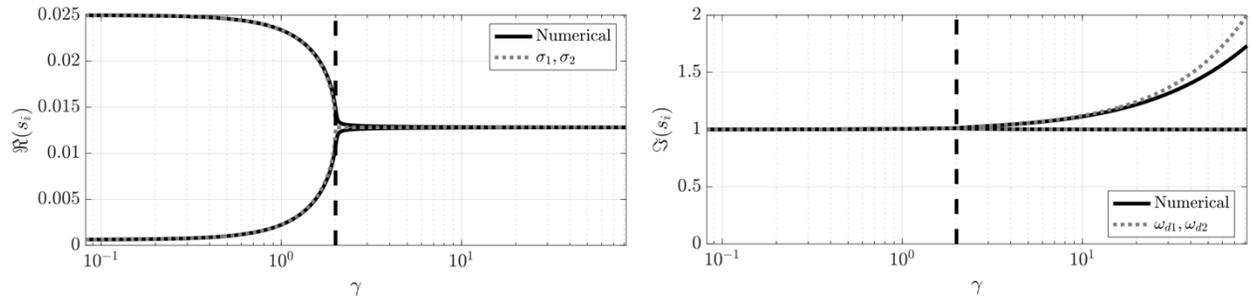

Figure 1. Dissipation rates (left) and fast frequencies (right) of the non-classically damped system with closely spaced modes of Fig. 1 (the critical value $\gamma_{cr} = 2$ is indicated by the vertical dashed line); note two distinct dissipation rates and single fast frequency close to unity for $\gamma < 2$, and two distinct fast frequencies close to unity and single dissipation rate for $2 < \gamma < 10$.

To validate the analytical predictions (5-6), we solved numerically the exact eigenvalue problem associated with system (1) by fixing the damping coefficients to $\lambda_1 = 0.00125, \lambda_2 = 0.05$, and varying the coupling stiffness $\beta$. Then we compared the real and imaginary parts of the numerically computed eigenvalues $s_i, i = 1,2$ to the analytically predicted dissipation rates and fast frequencies, respectively. These are depicted in Fig. 2, from which good agreement between the analytical predictions and direct numerical simulations is noted (at least up to moderate values of $\gamma$). In particular, the critical value $\gamma_{cr} = 2$ is confirmed, while the discrepancy in the fast frequencies observed for $\gamma > 10$ is due to the coupling not being small, and the two fast frequencies being well separated; as a result, the analytical prediction fails since it is based on the assumption of a single fast frequency close to unity.

In Figs. 3 and 4 we compare the results of direct numerical integrations of system (1) to the analytical CX-A predictions for the corresponding envelopes of the responses. Two types of impulsive excitations are considered: Case 1 corresponds to an impulse excitation of (normalized) intensity $\sqrt{2}$ applied to the heavily damped oscillator 2, while Case 2 corresponds to impulse excitation of intensity $\sqrt{2}$ applied to the lightly damped oscillator 1.

For $\gamma < 2$ (Fig. 3), we deduce that the amplitude of the directly forced oscillator 2 for Case 1 decays faster than that of the directly forced oscillator 1 for Case 2. This indicates that the decaying response of oscillator 2 is dominated by the higher dissipation rate ($\sigma_2$), while the response of oscillator 1 is dominated by the lower dissipation rate ($\sigma_1$); hence, the existence of two distinct dissipation rates is verified, as is the existence of a single fast frequency (the frequency of the oscillations inside the envelopes). The existence of a single fast frequency prevents the realization of recurrent beat phenomena in this case. We also note that reciprocity exists in the system since the response of the undriven oscillator is identical for each Case.

For $\gamma > 2$ (Fig. 4), however, we deduce the occurrence of recurrent beating phenomena irrespective of which of the two oscillators is directly excited, with energy being continuously exchanged between the two oscillators on a faster time scale. This makes sense since in this dynamical regime the responses possess *two distinct but closely-spaced fast frequencies*, $\omega_{d1}$ and $\omega_{d2}$, which give rise to the new *slow frequency in the envelope* equal to $\omega_d = (\omega_{d2} - \omega_{d1})/2$; at the same time, there is a single fast frequency of the carrier wave ($\omega_0$) which is the same to that for $\gamma < 2$. Moreover, no discernible change in the decaying behavior is observed for both oscillators, which demonstrates the existence of a single dissipation rate in the system. It follows that, in contrast to $\gamma < 2$, the envelopes of the responses of the two oscillators are decaying but also oscillatory. This gives rise to a (theoretically) infinite number of beat phenomena.

In addition, we Fourier-transformed the analytical solutions and computed the receptance frequency response functions (FRFs) to explore the response of the system in the context of modal analysis. The amplitude and phase of each of the computed receptance FRFs are shown at the right plots of Figs 3 and 4. For $\gamma < 2$ (Fig. 3) and considering Case 1 (excitation of the heavily damped oscillator 2), we deduce a single peak (resonance at the fast frequency $\omega_0 \approx 1$) in the FRF of the undriven oscillator 1, while there appears a dip around resonance in the FRF of the directly driven oscillator 2. This indicates intense modal interactions when the heavily damped oscillator 2 is driven. However, considering Case 2 (the lightly damped oscillator 1 is directly excited), the FRFs of both oscillators exhibit a single peak at resonance $\omega_0 \approx 1$, indicating the presence of a single fast frequency $\omega_0$, but, at the same time, *"concealing" the presence of two closely spaced modes in the system*. These results showcase the challenges that one faces when attempting to perform modal analysis of a non-classically damped system with closely spaced modes in this dynamical regime.

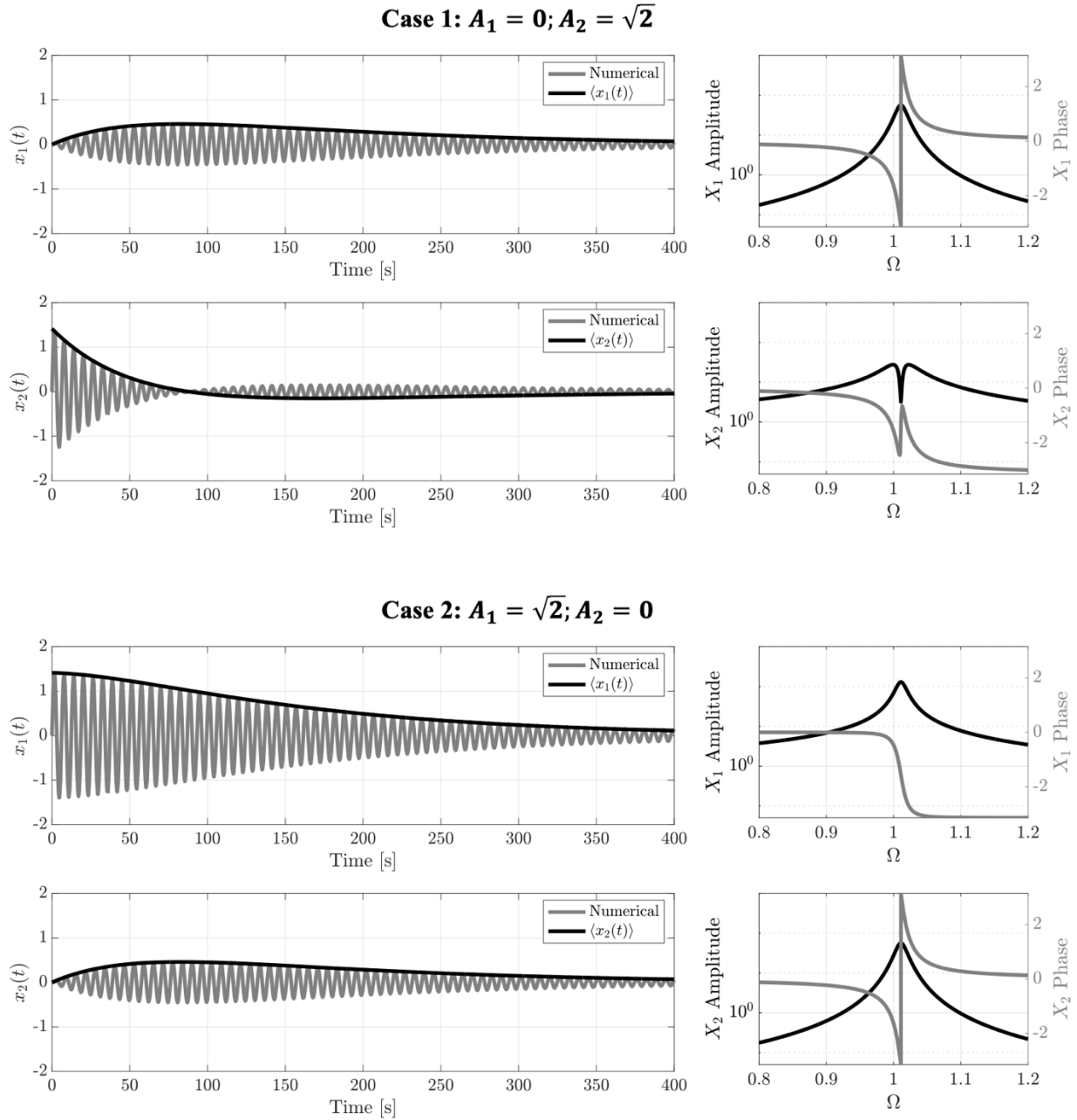

Figure 3. Comparison of numerical timeseries (gray lines) and analytically predicted envelopes of these time series (black lines) of the responses of system (1) with $\gamma = 1.8 < 2$: Upper plots correspond to Case 1 (impulsive excitation of only oscillator 2), and lower plots to Case 2 (impulse excitation of only oscillator 1); the corresponding Fourier transforms of these responses are shown on the right plots (modulus – black line, phase – gray line).

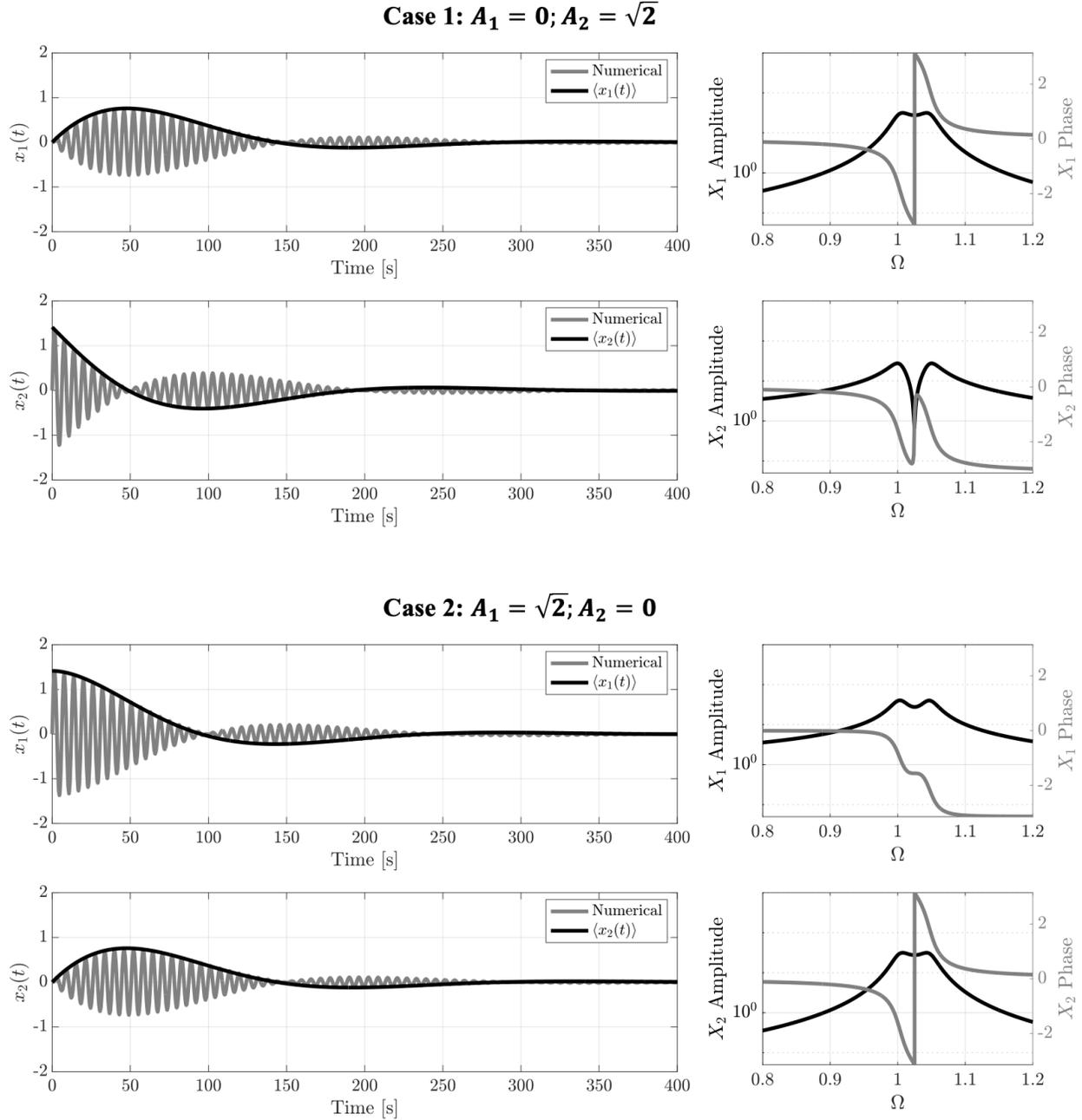

Figure 4. Comparison of numerical timeseries (gray lines) and analytically predicted envelopes of these time series (black lines) of the responses of system (1) with $\gamma = 4.1 > 2$: Upper plots correspond to Case 1 (impulsive excitation of only oscillator 2), and lower plots to Case 2 (impulse excitation of only oscillator 1); the corresponding Fourier transforms of these responses are shown on the right plots (modulus – black line, phase – gray line).

For $\gamma > 2$ (Fig. 4), all FRFs exhibit two peaks regardless of which oscillator is directly driven, indicating the presence of two fast (modal) frequencies $\omega_{d1}$ and $\omega_{d2}$, or, equivalently, two modes in the system. Lastly, for Case 1, again there is a dip in the FRF of oscillator 2, but in this case it occurs at the average $(\omega_{d1} + \omega_{d2})/2$ of the two fast frequencies. Again, these findings highlight the challenges to modal analysis in the frequency domain due to modal interactions.

## 3. Two distinct regimes of beat phenomena

To study in greater detail the beat phenomena that are generated by modal interactions, in this section we focus on the slow dynamics, i.e., the slowly decaying envelopes of the responses of the two oscillators predicted by the CX-A analysis (and verified by direct numerical simulations). Specifically, we consider the time instants $t_{mn,max}^{(k)}, m, n = 1,2, k = 1,2, ...$ where the slow envelopes in (5,6) reach either local maxima satisfying the condition, $\frac{d\langle x_{mn}(t)\rangle}{dt}\big|_{t_{mn,max}^{(k)}} = 0$ with $k$ being the order of the local maximum; or the time instants $t_{mn,0}^{(k)}$ where the envelopes vanish, i.e., $\langle x_{mn}(t_{i,0}^{(k)})\rangle = 0$. These results elucidate the formation of beat phenomena in the non-classically damped system with closely spaced modes, and their dependence on the parameter $\gamma$. This parameter dependence is important since, as shown previously, the critical value $\gamma_{cr} = 2$ divides two dynamical regimes, namely a regime where a single beat phenomenon is realized (for $\gamma < 2$, resulting in slower modal interaction) and a regime where an infinity of beat phenomena are formed (for $\gamma > 2$, resulting in faster modal interactions).

Considering the analytical approximations (5,6), first we consider the zeroes of the slow envelopes, which for $\gamma < 2$ are computed by:

$$\langle x_{12}(t)\rangle = 0 \Rightarrow (e^{-\sigma_1 t} - e^{-\sigma_2 t}) = 0 \tag{7a}$$

$$\langle x_{22}(t)\rangle = 0 \Rightarrow \frac{1}{2}(e^{-\sigma_1 t} + e^{-\sigma_2 t}) - \frac{1}{\sqrt{4-\gamma^2}}(e^{-\sigma_1 t} - e^{-\sigma_2 t}) = 0 \tag{7b}$$

$$\langle x_{11}(t)\rangle = 0 \Rightarrow e^{-\sigma_1 t} + e^{-\sigma_2 t} + \frac{2}{\sqrt{4-\gamma^2}}(e^{-\sigma_1 t} - e^{-\sigma_2 t}) = 0 \tag{7c}$$

Considering impulsive excitation of only one of the two oscillators (Case 1 for $A_1 = 0$, $A_2 = \sqrt{2}$, and Case 2 for $A_1 = \sqrt{2}$, $A_2 = 0$), it is clear that $\langle x_{12}(t)\rangle$ and $\langle x_{11}(t)\rangle$ cannot cross zero in finite time. This indicates that the envelope of the undriven oscillator cannot cross zero, so no repeated beat phenomena can occur for $\gamma < 2$. Rather, a single-cycle (single beat) energy exchange between the two oscillators takes place, yielding slow energy transfer during the modal interaction.

A different conclusion is obtained for $\gamma > 2$, where the zeros of the envelopes are computed according to:

$$\langle x_{12}(t)\rangle = 0 \Rightarrow \sin(\omega_d t) = 0 \Rightarrow t_{12,0}^{(k)} = \frac{k\pi}{\omega_d}, \quad k = 1,2, ...; \tag{8a}$$

$$\langle x_{22}(t)\rangle = 0 \Rightarrow \cos(\omega_d t) - \frac{2}{\sqrt{\gamma^2-4}}\sin(\omega_d t) = 0 \Rightarrow$$

$$t_{22,0}^{(k)} = \frac{1}{\omega_d}\left(\arctan\left(\frac{\sqrt{\gamma^2-4}}{2}\right) + (k-1)\pi\right), \quad k = 1,2, ... \tag{8b}$$

$$\langle x_{11}(t)\rangle = 0 \Rightarrow \cos(\omega_d t) + \frac{2}{\sqrt{\gamma^2-4}}\sin(\omega_d t) = 0 \Rightarrow \tag{8c}$$

$$t_{11,0}^{(k)} = \frac{1}{\omega_d}\left(\arctan\left(-\frac{\sqrt{\gamma^2-4}}{2}\right) + k\pi\right), \quad k = 1, 2, \ldots$$

Here we note an infinity of zeros for each slow envelope, indicating the existence of an infinite number of beat phenomena and repeated faster energy exchanges between oscillators (and modes).

Considering now the local maxima of the slow envelopes, again we consider the two different dynamical regimes which are defined depending on $\gamma$. For $\gamma < 2$, we obtain:

$$\frac{d\langle x_{12}(t)\rangle}{dt} = 0 \Rightarrow$$
$$-\sigma_1 e^{-\sigma_1 t} + \sigma_2 e^{-\sigma_2 t} = 0 \Rightarrow t_{12,max}^{(1)} = \frac{1}{\sigma_2 - \sigma_1}\log\left(\frac{\sigma_2}{\sigma_1}\right) > 0 \quad (9a)$$

$$\frac{d\langle x_{22}(t)\rangle}{dt} = 0 \Rightarrow$$
$$\left(-1 + \frac{2}{\sqrt{4-\gamma^2}}\right)\sigma_1 e^{-\sigma_1 t} - \left(1 + \frac{2}{\sqrt{4-\gamma^2}}\right)\sigma_2 e^{-\sigma_2 t} = 0 \Rightarrow \quad (9b)$$
$$t_{22,max}^{(1)} = \frac{1}{\sigma_2 - \sigma_1}\log\frac{\left(2+\sqrt{4-\gamma^2}\right)\sigma_2}{\left(2-\sqrt{4-\gamma^2}\right)\sigma_1}$$

$$\frac{d\langle x_{11}(t)\rangle}{dt} = 0 \Rightarrow$$
$$\left(1 + \frac{2}{\sqrt{4-\gamma^2}}\right)\sigma_1 e^{-\sigma_1 t} + \left(1 - \frac{2}{\sqrt{4-\gamma^2}}\right)\sigma_2 e^{-\sigma_2 t} = 0 \Rightarrow \quad (9c)$$
$$t_{11,max}^{(1)} = \frac{1}{\sigma_2 - \sigma_1}\log\frac{\left(2-\sqrt{4-\gamma^2}\right)\sigma_2}{\left(2+\sqrt{4-\gamma^2}\right)\sigma_1} < 0$$

From (9c), the computed time instant is always negative-valued, so the envelope of the lightly damped oscillator 1 will monotonically decrease when it is directly driven. In all other cases the envelopes of the two oscillators have a single maximum, which again confirms the absence of repeated beat phenomena in this case.

Lastly, for $\gamma > 2$ the local maxima of the envelopes are analytically approximated as follows:

$$\frac{d\langle x_{12}(t)\rangle}{dt} = 0 \Rightarrow$$
$$-\frac{\lambda_+}{2}\sin(\omega_d t) + \omega_d \cos(\omega_d t) = 0 \Rightarrow \quad (10a)$$
$$t_{12,max}^{(k)} = \frac{1}{\omega_D}\left(\arctan\left(\frac{2\omega_d}{\lambda_+}\right) + k\pi\right), k = 1, 2, \ldots$$

$$\frac{d\langle x_{22}(t)\rangle}{dt} = 0 \Rightarrow \quad (10b)$$

$$\left(-\frac{\lambda_+}{2} - \frac{2\omega_d}{\sqrt{\gamma^2 - 4}}\right)\cos(\omega_d t) + \left(\frac{\lambda_+}{\sqrt{\gamma^2 - 4}} - \omega_d\right)\sin(\omega_d t) = 0 \Rightarrow$$

$$t_{12,max}^{(k)} = \frac{1}{\omega_D}\left(\arctan\left(\frac{\lambda_+\sqrt{\gamma^2 - 4} + 4\omega_D}{2\lambda_+ - 2\omega_D\sqrt{\gamma^2 - 4}}\right) + (k-1)\pi\right), k = 1,2,\ldots$$

$$\frac{d\langle x_{11}(t)\rangle}{dt} = 0 \Rightarrow$$

$$\left(-\omega_d - \frac{\lambda_+}{\sqrt{\gamma^2 - 4}}\right)\sin(\omega_d t) - \left(\frac{\lambda_+}{2} - \frac{2\omega_d}{\sqrt{\gamma^2 - 4}}\right)\cos(\omega_d t) = 0 \Rightarrow \quad (10c)$$

$$t_{11,max}^{(k)} = \frac{1}{\omega_d}\left(\arctan\left(\frac{4\omega_d - \lambda_+\sqrt{\gamma^2 - 4}}{2\lambda_+ + 2\omega_d\sqrt{\gamma^2 - 4}}\right) + k\pi\right), k = 1,2,\ldots$$

Confirming the previous results, there is a countable infinity of local maxima of the envelopes of the two oscillators, i.e., a countable infinity of beat phenomena.

We depict now the first two local maxima and zeros of the envelopes of the responses of the two oscillators as functions of $\gamma$, for a system with $\lambda_1 = 0.00125$, $\lambda_2 = 0.05$, and varying coupling stiffness $\beta$. For comparison we examine both Case 1 – impulse excitation of intensity $\sqrt{2}$ applied to the heavily damped oscillator 2 (Fig. 5), and Case 2 – impulse excitation of intensity $\sqrt{2}$ applied to the lightly damped (right) oscillator 1 (Fig. 6).

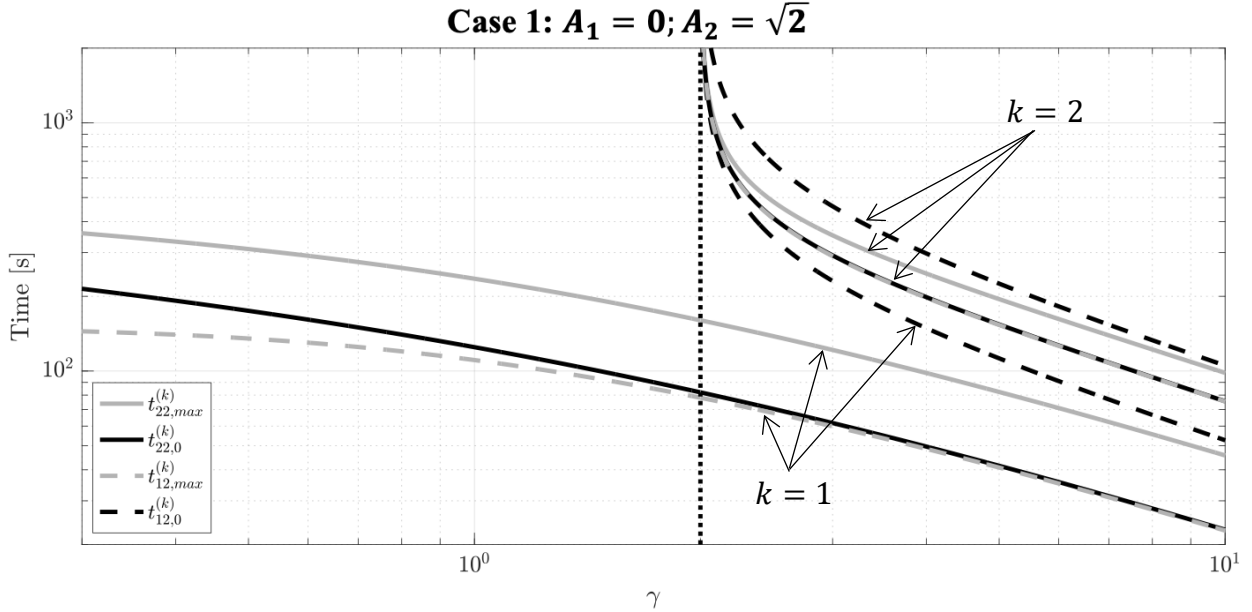

Figure 2. Leading-order local maxima and zeros of the envelopes of oscillators 1 and 2 as functions of $\gamma$: Case 1 (heavily damped oscillator 2 is directly excited); vertical dashed line indicates the critical value $\gamma_{cr} = 2$).

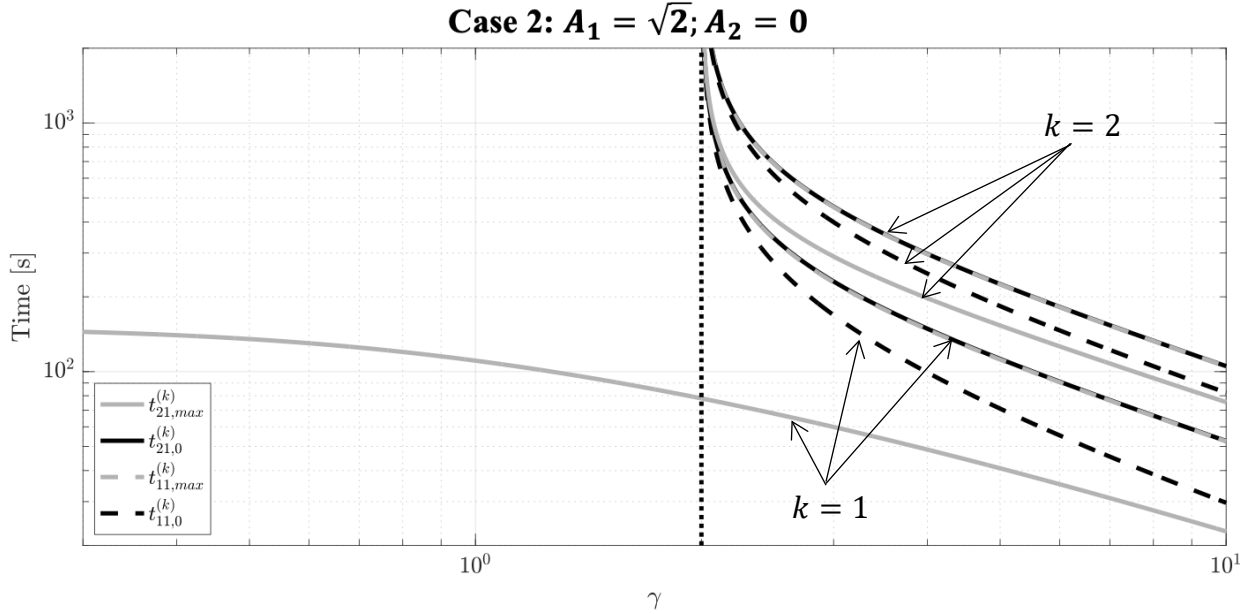

Figure 6. Leading-order local maxima and zeros of the envelopes of oscillators 1 and 2 as functions of $\gamma$: Case 2 (lightly damped oscillator 2 is directly excited); vertical dashed line indicates the critical value $\gamma_{cr} = 2$).

From these results the two distinct dynamical regimes are deduced for both cases of impulsive excitation. Namely, for $\gamma < 2$ a single beat phenomenon is realized, whereas for $\gamma > 2$ there is a sequence of beat phenomena – in Figs. 5 and 6 only the first two beat phenomena are depicted. Note that for Case 2 there exists no maximum in the envelope of the response of the directly excited oscillator 1 (since its response is monotonically decaying), in contrast to Case 1 where the envelope of the response of the directly excited oscillator 2 has a single local maximum as well as a zero (the two corresponding time instants converge to each other and become identical at $\gamma_{cr} = 2$). In addition, we note that the phase differences between successive maxima and zeros of the envelopes of the two oscillators are approximately equal to $\sim \pi/2$. This indicates that energy is continuously exchanged at a relatively fast scale (compared to $\gamma < 2$) between the two oscillators, approximately every period $T = \pi/\omega_d$.

## 4. Experimental validation

The experimental fixture shown in Fig. 7 was constructed to validate the two theoretically predicted dynamical regimes. It consists of two oscillators in the form of rigid masses (blocks) grounded by flexures, coupled with a thin steel flexure. Each oscillator is constructed using an aluminum block with length of 0.152 m, width of 0.152 m, and thickness of 0.0127 m. Each oscillator includes a protruding shaft with a threaded rod, which is used to connect them together with a thin steel flexure. The oscillators are installed on an optical table using aluminum L-brackets and a pair of thin steel flexures with thicknesses of 0.889 mm and 0.762 mm for oscillator 1 and oscillator 2, respectively. Additionally, oscillator 2 includes a polyurethane/polyethylene foam block inside it, which introduces additional stiffness and additional damping. We compensate for this additional stiffness by using thicker flexures on oscillator 1 compared to oscillator 2. The masses of the oscillators are identical, equaling 1.031 kg, which includes all bolts, washers, and the shaft but does not include the foam block. The foam block has a mass of 0.063 kg, bringing the total mass of oscillator 2 up to $m_2 = 1.094$ kg; accordingly, an additional aluminum block of mass 0.042 kg was added to oscillator 1 to compensate for this discrepancy, which increases its total mass to $m_1 = 1.073$ kg. The stiffness and mass compensation was implemented to ensure that the two oscillators had the same natural frequency.

A pair of automatic modal hammers (Singh and Moore, 2022; Nasr et al., 2024) were used to excite each oscillator and the resulting response was measured using high-sensitivity accelerometers (PCB Piezotronics model 333B50) with nominal sensitivities of 102 mV/(m/s$^2$). The modal hammer and accelerometers were measured using a HBK QuantumX MX1601B module with catman® software, then exported and analyzed using MATLAB®. The response was measured for 30.5 seconds at a rate of 19,200 Hz with a pretrigger of 0.5 seconds. The resulting accelerations were numerically integrated, then high-pass filtered to obtain the velocities. This process was repeated with the velocities to obtain the displacements. The integration was performed using the *cumtrapz* function in MATLAB® and the filter used was a third-order Butterworth filter with a cutoff frequency of 10 Hz.

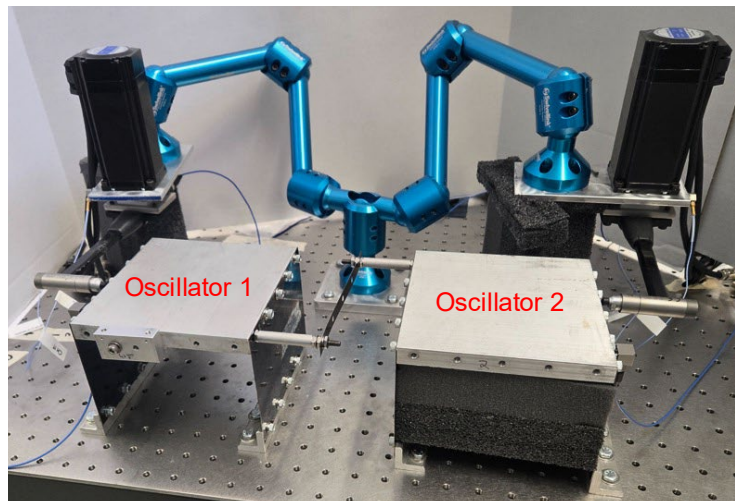

Figure 7. The experimental system consisting of a pair of near-identical harmonic oscillators (but for the damping) coupled with a thin steel flexure and excited by a pair of automatic modal hammers.

The oscillators were first tested in an uncoupled state to determine their individual dynamics. In these tests, a large steel mass was placed on the undriven oscillator to significantly decrease its natural frequency so to avoid interactions between the two oscillators. Impulses of 65.7 N and 64.8 N were applied to oscillators 1 and 2, respectively, to excite them in a linear regime (small displacement). The resulting displacements and corresponding FRFs are shown in Fig. 8, which shows that two oscillators have nearly identical natural frequencies, but oscillator 2 has a significantly higher damping. The natural frequency and damping ratio for each oscillator were identified using the rational polynomial fraction method (Richardson and Formenti, 1982). The natural frequency of both oscillators is 26.62 Hz, while the viscous damping ratios are estimated as, 0.0017 and 0.0451 for oscillators 1 and 2, respectively. These values result in a stiffness of $k_1 = 30019$ N/m and damping coefficient of $c_1 = 0.614$ Ns/m for oscillator 1 and a stiffness of $k_2 = 30606$ N/m and damping coefficient of $c_2 = 16.51$ Ns/m for oscillator 2. The equivalent non-dimensional stiffness and damping for oscillator 1 are $\beta_1 = 1$ and $\lambda_1 = 0.003$, respectively, whereas for oscillator 2, $\beta_2 = 1.02$ and $\lambda_2 = 0.092$, respectively. Thus, we find that the experimental system is nearly identical to the theoretical system (1), and the small difference in the stiffnesses is negligible.

Next, the oscillators were coupled together using two different steel flexures with length 0.152 m, width of 0.0127 m, and thicknesses of 0.254 mm and 0.508 mm. Two different flexures were used to produce responses for $\gamma < 2$ and $\gamma > 2$, as required in the previous analysis. For each flexure, we measured the response of the system for excitation applied to each oscillator to reproduce the two forcing Cases 1 and 2 considered in the theoretical study. In all measurements, the applied impulsive forces had nominal amplitude of 60 N, ensuring that linear responses were excited in both Cases. The stiffness of each flexure is determined using the second natural frequency, $\omega_2$, because the first natural frequency is the same as that of the uncoupled oscillators. Leveraging the eigenvalue problem, we solve for the coupling stiffness as a function of the masses, stiffnesses, and second natural frequency as,

$$K = \frac{-k_1 k_2 + (k_2 m_1 + k_1 m_2)\omega_2^2 - m_1 m_2 \omega_2^4}{k_1 + k_2 - (m_1 + m_2)\omega_2^2}, \qquad (11)$$

where $K$ is the coupling stiffness. The measured second natural frequencies are 27.64 Hz and 28.74 Hz for the 0.254-mm-thick flexure and the 0.508-mm-thick flexure, respectively, resulting in stiffnesses of $K = 1181.4$ N/m and $K = 2503$ N/m, respectively. Hence, the non-dimensional coupling stiffnesses are $\beta = 0.039$ and $\beta = 0.083$ for the 0.254-mm-thick flexure and the 0.508-mm-thick flexure, respectively.

Figure 8. The measured displacement responses and corresponding receptance FRFs for each oscillator measured in the uncoupled state.

We leverage a slightly different, but equivalent formulation for $\gamma$,

$$\gamma = \frac{4K}{\omega_n(c_2 - c_1)}, \qquad (12)$$

where $\omega_n$ is the natural frequency of each oscillator in the uncoupled state and is also equal to the first natural frequency of the coupled system. Equation (12) allows for computing $\gamma$ using the experimental parameters directly and results in $\gamma = 1.78$ and $\gamma = 3.77$ for the 0.254-mm-thick flexure and the 0.508-mm-thick flexure, respectively.

For both Cases of impulsive excitations and for $\gamma = 1.78$ we present the experimental measurements in Fig. 9a, where we compare the displacement response of the driven oscillator in the top panel and that for the undriven oscillator in the bottom panel. The top panel clearly shows that oscillator 2 is dominated by the higher dissipation rate while oscillator 1 is dominated by the lower dissipation rate. This indicates the presence of two distinct dissipation rates in the responses but a single fast frequency. Hence, there is absence of recurring beat phenomena, as predicted by the analysis. The bottom panel shows that reciprocity is maintained in the system just as in the theoretical system. The corresponding results for $\gamma = 3.77$ are depicted in Fig. 10, where the top panel shows the displacement of the driven oscillator in each Case while the bottom panel presents the displacement of the undriven oscillator. The displacements of the two oscillators reveal the presence of a slow frequency in the variations of the envelopes (in addition to the fast frequency of the oscillations inside the envelopes), which gives rise to recurring beat phenomena; while the bottom panel shows that reciprocity is satisfied in the experimental system. These results exactly match the theoretical predictions and experimentally confirm the role that $\gamma$ plays in controlling the generation of beat phenomena and modal interactions in the non-classically damped system with closely spaced modes.

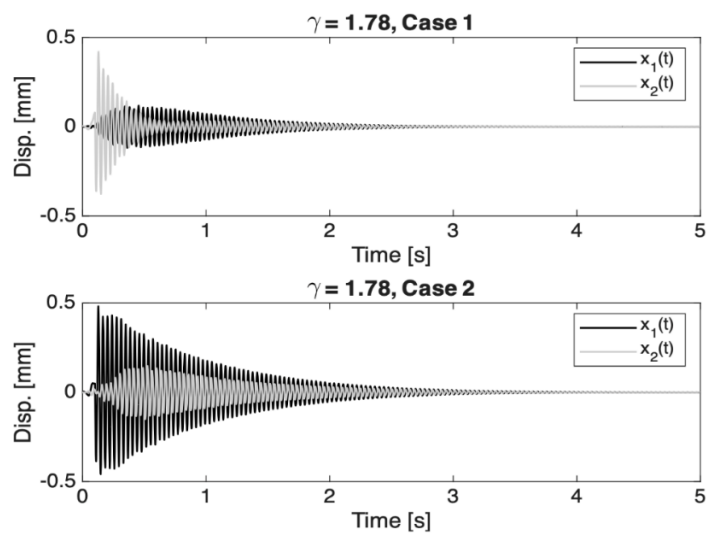

Figure 9. Experimental displacements for impulsive excitation Cases 1 and 2 for the system with $\gamma = 1.78 < 2$; note absence of recurring beat phenomena.

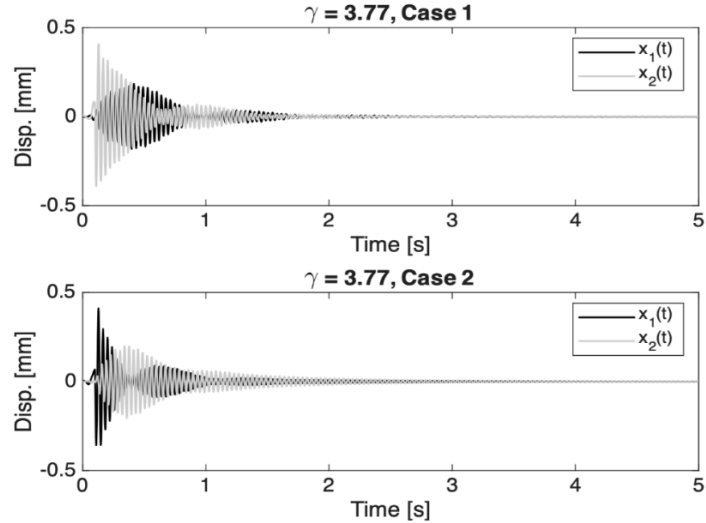

Figure 10. Experimental displacements for impulsive excitation Cases 1 and 2 for the system with $\gamma = 3.77 > 2$; note the realization of recurring beat phenomena.

## 5. Concluding remarks

The combined effects of a non-proportional damping distribution and proximity of vibration modes on the dynamics of a system of coupled oscillators were analyzed in this work. It was shown that a critical parameter in the form of coupling to damping non-proportionality ratio governs the modal interactions in this system: When this ratio is below a critical value the interaction between the closely spaced modes occurs over a relatively slow time scale and no recurring beat phenomena can occur; this happened when the ratio exceeds the critical value, giving rise to recurring beat phenomena and modal interactions over a faster time scale. The physical explanation for these dynamics is the existence of two distinct dissipation rates and a single resonance (fast) frequency in the former dynamical regime, and of two distinct resonance frequencies and a single dissipation rate in the latter. The theoretical predictions were fully verified experimentally. We believe these results provide insight into the well-known challenges that one faces when performing modal analysis in the frequency domain of non-classically damped coupled oscillators with closely spaced modes. The findings of this work provide a new framework for understanding and modeling modal interactions in a broad class of non-classically damped systems with closely spaced modes, including systems with families of interacting modes, or systems incorporating weak coupling and mechanical joints introducing nonlinearities such as hysteresis or clearance. Moreover, the results of this work can benefit experimental modal analysis of structures with closely spaced modes, a field that is often encountered in vibration engineering practice.